**Title:**

Watching charge separation in nanoantennas by ultrafast point-projection electron microscopy

**Authors**:

Jan Vogelsang[1]*, Germann Hergert[1], Dong Wang[2], Petra Groß[1], Christoph Lienau[1,3]*

**Affiliations**:

[1] Carl von Ossietzky Universität, Institut für Physik and Center of Interface Science, 26129 Oldenburg, Niedersachsen, Germany.

[2] TU Ilmenau, Institut für Werkstofftechnik und Institut für Mikro- und Nanotechnologien, 98693 Ilmenau, Thüringen, Germany.

[3] Carl von Ossietzky Universität, Forschungszentrum Neurosensorik, 26129 Oldenburg, Niedersachsen, Germany.

* e-mail: jan.vogelsang@uni-oldenburg.de; christoph.lienau@uni-oldenburg.de

**Introductory paragraph:**

Watching the motion of electrons on their natural nanometre length- and femtosecond time scales is a fundamental goal and an open challenge of contemporary ultrafast science[1-5]. Optical techniques and electron microscopy currently mostly provide either ultrahigh temporal or spatial resolution, yet, microscopy techniques with combined space-time resolution need further development[6-11]. Here we create an ultrafast electron source by plasmon nanofocusing on a sharp gold taper and implement this source in an ultrafast point-projection electron microscope. This source is used, in an optical pump – electron probe experiment, to study ultrafast photoemission from a nanometer-sized plasmonic antenna.[12-15] We show that the real space motion of the photoemitted electrons and residual holes in the metal is probed with 20-nm spatial resolution and 25-fs time resolution. This is a step forward towards time-resolved microscopy of electronic motion in nanostructures.

**Main Text:**

The light-induced separation of charge carriers is one of the most fundamental processes in nature. It forms the basis for a vast class of electron transfer reactions in donor-acceptor or light-harvesting complexes[1,4,5] as well as for a multitude of technological applications, e.g., in photocathodes[12,16], -diodes[6,17] and solar cells[3,18,19]. In recent years, nanostructures are becoming more and more important for enhancing charge separation, for example in photovoltaic devices[20,21], and in particular also in higher harmonic generation from solids[22] and ultrafast electron microscopy (UEM)[8-11,23,24]. For UEM, e.g., metallic nanotips, driven by strongly enhanced local optical fields, emerge as a new and versatile class of nanoscale electron sources[12-15,25]. In all these structures, the local, light-induced birth of charge carriers is intimately connected to an ultrafast real-space motion of the photo-generated electron and hole wave packets.

These transport phenomena typically occur on ~10-fs time and ~10-nm length scales. As such, their direct visualization inherently requires ultrafast microscopy techniques with nanometre resolution. Despite recent progress in developing such methods[2,10,11,23,26,27], the required spatio-temporal resolution and measurement sensitivity is still challenging to obtain. Ultrafast optical techniques provide attosecond time resolution but are inherently diffraction-limited. High-photon energy XUV or X-ray spectroscopies can in principle improve this resolution yet lack the sensitivity to probe dynamics in single nanostructures. In contrast, time-resolved electron microscopies can reach few-nm resolution while so far being limited to 100s of fs-time resolution[28]. More specifically, point-projection microscopes feature shorter propagation distances than conventional electron microscopes, but are currently still limited to time resolutions of 100 fs or more by dispersion[11,29]. Fs-photoelectron emission microscopy (PEEM) has successfully been used to image local electric fields at surfaces with few tens of fs and few tens of nm resolution[27,30,31]. PEEM, however, necessarily requires high bias fields, inherently disturbing the free motion of photo-released charge carriers. Here, we make use of nanofocusing of surface plasmons on a conical gold taper to emit ultrashort photoelectron pulses from the nanometer-sized apex of the taper. This source is implemented in a point-projection electron microscope (UPEM) and provides a combined spatio-temporal resolution of 20 nm and 25 fs, respectively. We use this microscope to directly track the motion of electrons that are photo-released from the hot spot of a single plasmonic nanoantenna and see how they separate from the positive charges that are left behind in the metal.

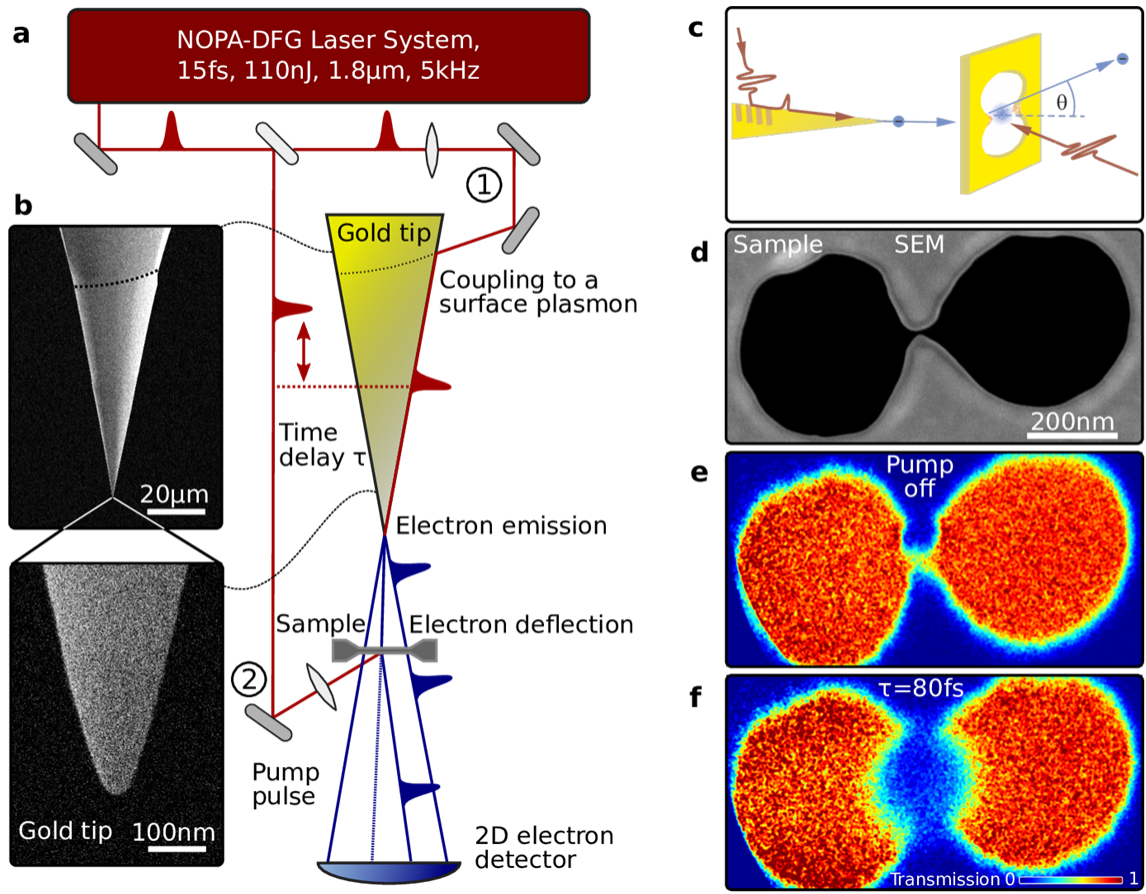

**Figure 1 | Ultrafast point-projection electron microscopy (UPEM). a**, Schematic UPEM setup. **b**, Scanning electron microscope (SEM) images of the gold taper used as the electron emitter. **c**, UPEM imaging of plasmon-enhanced photoemission. Ultrashort electron probe pulses generated by plasmonic nanofocusing are deflected off a cloud of electrons photo-released from the gap of a gold nanoantenna. **d**, SEM image of a double-nanohole antenna in a 30-nm thick free-standing gold film. **e**, Point-projection image recorded in the absence of a pump laser, mapping the shape of the double-hole nanoantenna. **f**, The transient point-projection image recorded 80 fs after illuminating the sample with a femtosecond laser pulse is drastically different: The photo-released electrons cause a local reduction in probe electron transmission in the region around the antenna gap. (123 words)

In our microscope, we generate ultrashort electron pulses by nanofocusing[15,32-35] of femtosecond surface plasmon polariton (SPP) pulses on a sharp gold nanotaper (Fig. 1a). Few-cycle laser pulses at a wavelength of 1.8 μm and with a duration of 15 fs[36] are focused onto the taper shaft, launching SPPs via a grain boundary at a distance of 80 μm from the taper apex acting as localized electron emitter. At this wavelength, SPP losses are low, resulting in long-distance plasmon propagation and nanofocusing to a ~ 15 nm-sized focus at the very apex of the monocrystalline gold taper (Fig. 1b)[37]. This nanofocusing is so efficient that high local SPP fields with amplitudes of up to 10 V/nm are generated. These are sufficiently high to release about one electron per pulse from a sub-10-nm apex region in a fifth-order photoemission process. The high nonlinearity of this emission efficiently restricts photoemission to the very apex region and effectively creates a free-standing nanometre-sized electron source with sub-10-fs pulse duration[15]. A characterization of the time structure of the SPP field is provided in the Supplementary Information.

This free-standing electron source delivers the probe pulses in our UPEM. In earlier implementations of time-resolved point-projection microscopes[11,38,39], direct illumination of the laser apex has been used to trigger photoemission. The intense, diffraction limited laser spot used for photoemission has typical diameters of a few microns and thus undesired excitation of the sample can only be prevented by limiting the emitter-sample separation to at least a few (tens of) microns. This inherently restricts both the spatial (~ 100 nm)[11,35,38] and temporal resolution (~ 100 fs)[11,39] of point-projection microscopy. In contrast, the nanofocused electron source uses evanescent SPP fields to drive photoemission and thus provides the critical advantage of permitting ultrasmall emitter-sample separations.

In our UPEM, the emitted electrons are accelerated towards the sample by a 60 V bias, reducing their relative kinetic energy spread. The incident, divergent electron beam is diffracted off the sample. An image of the interference of transmitted and diffracted waves, magnified by the ratio between detector-emitter (75 mm) and sample-emitter (2700 nm) distance, is recorded on a microchannel plate detector. In static experiments, a similar design already resulted in sub-nm resolution holographic imaging[40]. Under our conditions, the de Broglie wavelength of the incident electrons (~ 0.15 nm) is smaller than the sample thickness and the image can be explained in the ray tracing limit[10]. The short emitter-sample distance allows us to operate the microscope at high magnification (30.000x) and -- most importantly -- effectively suppresses any temporal spreading of the electrons prior to the interaction with the sample. This is the key for advancing the time resolution of electron microscopy in the present experiment.

Here, we use this microscope to study the ultrafast dynamics of photoemitted electrons from a single plasmonic nanoantenna in an experimental configuration that is schematically depicted in Fig. 1c. We design a plasmonic nanoresonator by milling two adjacent holes with a hole diameter of 400 nm in a 30-nm-thin free-standing polycrystalline gold film (Fig. 1d). A small, ~30 nm wide channel connecting the two holes transforms the structure into a nanogap antenna. We make use of the field enhancement in the gap region to induce localized electron photoemission. For this, we illuminate the back side of the antenna with a second time-shifted replica of the 15-fs laser pulse at 1.8 μm. For linearly polarized excitation along the antenna arms, photoemission is induced at a peak electric field strength of 0.6 V/nm. In our experiments, these electrons cannot reach the detector since they are blocked by a 40 V sample-detector bias. The field amplitude of the pump laser is so weak that it does not induce photoemission from the taper apex.

This now allows us to measure background-free point-projection images of the nanoantenna. Hence, any apparent changes of the transient UPEM images are a direct consequence of the interaction of the probing electrons with the optically excited nanoantenna. In the absence of a pump laser, the UPEM image reveals a spatially homogeneous transmission of the probe electrons through the transparent regions of the antenna (Fig. 1e). For a fixed time delay of 80 fs between optical pump and electron probe, the transmission is largely blocked in a sharply confined region around the channel gap (Fig. 1f). This blocking can be understood as the deflection of the electron probe beam by the cloud of low-energy electrons that is photoemitted from the gap of the antenna[29,41,42]. By changing the time delay between optical pump and electron probe we can now create a movie of how this charge cloud evolves in space and time.

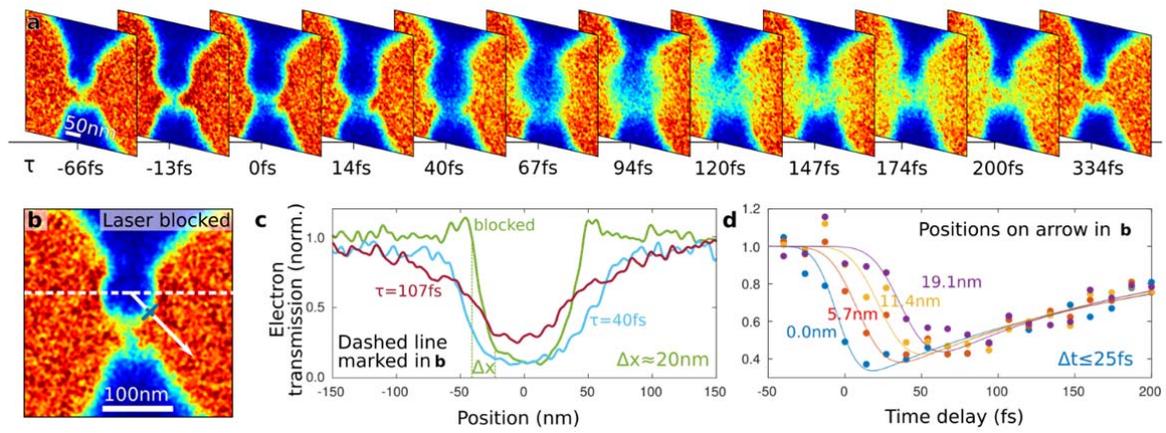

**Figure 2 | Dynamics of photoemission from the gap of a single plasmonic nanoantenna. a**, Series of transient UPEM images recorded for different time delays $\tau$ between the laser excitation pulses and probe electrons. UPEM images of the central gap region are displayed and the electron transmission is color-coded from 0 to 1 using the colour bar shown in Fig. 1f. The photoelectrons propagate away from the antenna gap, resulting in a transient, local reduction in electron transmission vanishing within 200 fs. **b**, UPEM image recorded without laser excitation. **c**, Cross cut along the dashed white line shown in B at different delay times, giving a spatial resolution of 20 nm. **d**, Transmission signal as a function of time at four equidistant positions along the white arrow in b. The electron signal decreases within 25 fs, the upper limit of the temporal resolution. (139 words)

Figure 2a shows a time sequence of UPEM images of the relevant gap region. The zero of time, $\tau = 0$ fs, denotes coincidence of optical and electron pulse maxima in the sample plane. For negative time delays, the geometric shape of the nanoantenna is imaged with a resolution of about 20 nm, as in the case of a blocked pump laser (Fig. 2b). Around time zero, a reduction of transmission in the central gap region sets in and the image becomes slightly blurred (Fig. 2c). Quickly, an almost circular blocking region emerges around the antenna gap. Its diameter expands in time and reaches a value of 200 nm at 80 fs. For longer time delays, the diameter further increases, but now the rim of the blocking region washes out and the central region becomes partially transparent again. For $\tau > 150$ fs, blurred images of the nanoresonator re-emerge until, at $\tau > 300$ fs, the images are virtually indistinguishable from those recorded without pump. For the currently reached spatial resolution of 20°nm, we estimate an electric field sensitivity of a few $10^7$ V/m, as discussed in the Supplementary Information.

We determine the temporal resolution of our microscope by evaluating the change in transmission at a position close to the upper rim of the gap antenna, marked with a blue tick on the white arrow in Fig. 2b. The transmission dynamics are shown in Fig. 2d (blue circles) and reveal a decrease in transmission within 25 fs (10%-90% criterion). This puts an upper bound on the time resolution of our UPEM. The reduction in electron transmission vanishes on a 100-fs time scale. The effect of the distance between gap antenna and electron probe on the dynamics is shown by the additional curves in Fig. 2d. A 6-nm increase in gap-probe distance (red circles) results in a sizeable time shift of the onset of electron deflection by ~ 11 fs. For time delays of <11 fs, the probe electrons at this spatial position do not yet see those electrons that are released from the antenna rim at earlier times. This shows that, already at this small gap-probe distance, photoemission from the gap antenna does not immediately result in a deflection of the probe beam. Instead, this requires a finite propagation of the released electrons from the rim of the gap to the probe position. Two conclusions can immediately be drawn. First, it appears that the deflection of the probe electrons is only sensitive to charges within less than 15-nm distance, since the probe electrons are not affected by those electrons that are emitted near the aperture rim. Second, the ratio between time shift and gap-probe distance provides a direct measure for the speed of propagation of the fastest electrons in the photo-released cloud. We estimate a speed of 0.5 nm/fs or $c_0/600$[42]. This corresponds to a kinetic energy of 0.7 eV, close to our photon energy. Upon further increase in gap-probe distance, the observed time delay increases linearly, confirming the picture of a ballistic propagation of the fastest electrons at $c_0/600$. In such a ballistic transport picture, we would expect that the dip in transmission vanishes as soon as the electron cloud has moved out of the probe volume. The finite persistence of the dip for ~100 fs therefore points to a broad distribution of kinetic energies, i.e., propagation speeds, of the released electrons.

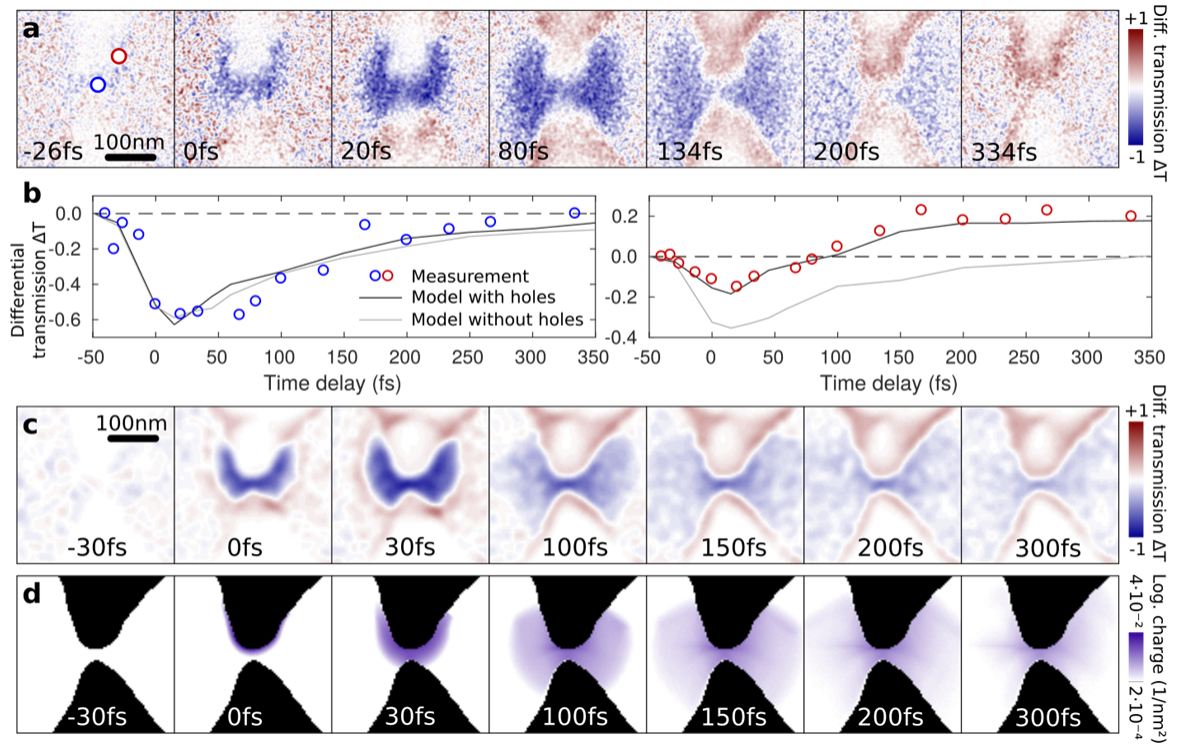

**Figure 3 | Differential UPEM transmission images and electron trajectory simulations. a**, Series of differential UPEM images, created by subtracting a background image recorded at $\tau = -66$ fs from transient UPEM images similar to those in Fig. 2a, but recorded for a pump field strength of 0.7 V/nm. A transient reduction in electron transmission is color-coded in blue and is dominant in the gap region. A transient increase in transmission (red) appears at later delay times in the two vertical arms of the antenna. Here, electrons are deflected into a non-transparent region due to residual positive charges on the antenna arms. The charging is more pronounced in the upper antenna arm acting as electron source. **b**, Temporal evolution of the differential transmission at two positions marked in a (open circles), together with the simulated evolution derived from a model without (grey curves) and with holes (black curves). **c**, Simulated differential UPEM images revealing both the expanding shadow in the gap region as well as the transmission enhancement in the antenna arms. **d**, Transient evolution of the charge density deduced from the trajectory simulations. (168 words)

To analyse the photoemission process in further detail, we present difference images between the electron transmission at finite time delays and that at a delay of -66fs, recorded with probe electrons arriving well before the laser pump (Fig. 3a). These data have been taken at slightly increased pump field strengths of 0.7 V/nm. The images show, color-coded in blue, the space-time dynamics of the drop in transmission due to photoemission from the nanoantenna. The spreading of the released electron cloud and its vanishing within 300 fs are evident. The dynamics of the differential transmission $\Delta T$, at a fixed position (blue circle) in the transparent region close to the antenna gap is shown in Fig. 3b. Interestingly, near the antenna rim, the electron transmission becomes larger than in the absence of the pump, in contrast to the signal drop created by the released electron cloud. This signal increase persists for delays beyond our measurement range, as is shown exemplarily in Fig. 3b (red circles).

We have performed classical electron trajectory simulations to understand these dynamics. In these simulations, we model the probe electrons as a beam of single, point-like charges that are deflected by the Coulomb fields generated by a randomly distributed cloud of photoelectrons. These electrons are placed at the surface of the nanoantenna, in a 50-nm wide region matching the local surface plasmon mode profile of the gap antenna. The electrons are created within a time window set by the pump laser and the kinetic energies are randomly chosen. For simplicity, we assumed a uniform distribution of their velocities up to a maximum set by the photon energy. Simulations of the resulting differential transmission images are seen in Fig. 3c. When choosing an appropriate number of released electrons, approximately 30 per pulse, the space-time dynamics of the experimentally observed drop in differential transmission is nicely reproduced. This strongly supports that the deflection of the probe electrons quantitatively maps the expansion of the photo-released electron cloud, further substantiated by earlier works which have studied similar phenomena albeit with lower spatio-temporal resolution[29,41-43]. The space-time dynamics of the electrons released from the antenna rim that is predicted by our model simulations is shown in Fig. 3d. Evidently, these simulations account well for the differential transmission dynamics in the transparent region of the nanoantenna. Yet, we cannot reproduce the persistent increase in differential transmission that is seen in Fig. 3a for probe positions near the antenna rim if we restrict the simulations only to a light-driven release of photoelectrons. Instead, we are led to assume that photoemission results in a build-up of positive charges at the metal surface near the apices of the two antenna arms. We have added an appropriate number of positive charges on each of the arms in our simulations. As is evident from Fig. 3c, this leads to a deflection of probe electrons into the otherwise obscured, non-transparent regions in the outer rim of the antenna arms. Directly at the edge of the antenna, the finite spatial resolution of our electron probe leads, initially, to a decrease in transmission due to the release of photoelectrons while, at later times, the transmission enhancement due to positive charge deflection dominates. Representative $\Delta T$ dynamics are shown in Fig. 3b (red circles). This transition between negative and positive $\Delta T$ can only be understood by assuming that the photoemission results in a persistent, positive charging of the upper arm of metal antenna. In principle, the observed charging may either be accounted for by photo-induced holes at the inside of the metal or by positively charged, long-lived surface states. Conceptually similar studies of photoelectron deflection by charge-separated electric fields have been performed earlier, with picosecond temporal and tens of microns spatial resolution, e.g., on cluster plasmas[40,43], copper films surfaces or near graphite surfaces[29]. Our UPEM techniques advances the space-time resolution of such deflection studies to the 10 nm / 10 fs regime, opening up exciting avenues for probing photoinduced charge transfer and separation dynamics in individual nanostructures[35] with a time resolution that is sufficient, for instance, for probing the effects of electron-phonon interactions on those dynamics[3].

This substantial improvement in space-time resolution has been accomplished by implementing plasmonically-enhanced multi-photon photoemission from sharp metal tapers, creating a freestanding source of ultrafast photoelectron pulses. In a proof-of-principle experiment, we have used this source here to study the ultrafast release and expansion of a cloud of charges from a single nanometre-sized plasmonic antenna providing a direct visualization of electron dynamics and charge separation in nanostructures on ultrafast time scales. In future applications, it seems most interesting to combine this electron source with ultrafast phase-resolved streaking schemes to coherently control low energy photoelectron deflection by single nanostructures. In particular, for small nanostructures, for which the transit time of the electrons through their optical near field is less than half an optical cycle, this promises imaging of coherent electrodynamic fields near surfaces with nanometre spatial and sub-cycle temporal resolution and may be a key for probing local and ultrafast charge carrier dynamics in nanostructures by deflecting passing electrons[11]. Being an intrinsically coherent diffraction scheme, we expect it, in the near future, to enable key advances in transient electron holography, potentially providing new approaches for gaining three-dimensional transient electronic information from individual nanosystems.

**Acknowledgments:** We thank P. Schaaf for support in the sample fabrication, and H.-W. Fink and J.-N. Longchamp for helpful discussions. Simulations were performed at the HPC Cluster CARL in Oldenburg (DFG INST 184/157-1 FUGG). We acknowledge financial support by the Deutsche Forschungsgemeinschaft (SPP1391, SPP1839, SPP1840), the German-Israeli Foundation (GIF grant no. 1256), and the Korea Foundation for International Cooperation of Science and Technology (Global Research Laboratory project, K20815000003). J.V. acknowledges a personal grant from the Studienstiftung des Deutschen Volkes.

**Author Contributions:** C.L. initiated the project. J.V. implemented the set-up. D.W. fabricated the gold film. J.V. and G.H. carried out the experiments. J.V., P.G., and C.L. analysed and interpreted the data. J.V. performed the numerical trajectory calculations. J.V., P.G., and C.L. prepared the manuscript. All authors contributed to the final version of the manuscript.

**Competing Financial Interests statement:** The authors declare no competing financial interests.

**References:**

1.   Scholes, G. D. *et al.* Using coherence to enhance function in chemical and biophysical systems. *Nature* **543**, 647 (2017).

2. Cocker, T. L., Peller, D., Yu, P., Repp, J., Huber, R. Tracking the ultrafast motion of a single molecule by femtosecond orbital imaging. *Nature* **539**, 263 (2016).

3. Falke, S. M. *et al.* Coherent ultrafast charge transfer in an organic photovoltaic blend. *Science* **344**, 1001 (2014).

4. Rozzi, C. A. *et al.* Quantum coherence controls the charge separation in a prototypical artificial light-harvesting system. *Nature Communications* **4**, 1602 (2013).

5. Calegari, F. *et al.* Ultrafast electron dynamics in phenylalanine initiated by attosecond pulses. *Science* **346**, 336 (2014).

6. Schiffrin, A. *et al.* Optical-field-induced current in dielectrics. *Nature* **493**, 70 (2013).

7. Sommer, A. *et al.* Attosecond nonlinear polarization and light–matter energy transfer in solids. *Nature* **534**, 86 (2016).

8. Zewail, A. H. Four-dimensional electron microscopy. *Science* **328**, 187 (2010).

9. Barwick, B., Zewail, A. H. Photonics and plasmonics in 4D ultrafast electron microscopy. *ACS Photonics* **2**, 1391 (2015).

10. Ryabov, A., Baum, P. Electron microscopy of electromagnetic waveforms. *Science* **353**, 374 (2016).

11. Müller, M., Paarmann, A., Ernstorfer, R. Femtosecond electrons probing currents and atomic structure in nanomaterials. *Nature Communications* **5**, 5292 (2014).

12. Herink, G., Solli, D., Gulde, M., Ropers, C. Field-driven photoemission from nanostructures quenches the quiver motion. *Nature* **483**, 190 (2012).

13. Hommelhoff, P., Sortais, Y., Aghajani-Talesh, A., Kasevich, M. A. Field emission tip as a nanometer source of free electron femtosecond pulses. *Physical Review Letters* **96**, 077401 (2006).

14. Ropers, C., Solli, D., Schulz, C., Lienau, C., Elsaesser, T. Localized multiphoton emission of femtosecond electron pulses from metal nanotips. *Physical Review Letters* **98**, 043907 (2007).

15. Vogelsang, J. *et al.* Ultrafast electron emission from a sharp metal nanotaper driven by adiabatic nanofocusing of surface plasmons. *Nano Letters* **15**, 4685 (2015).

16. Swanwick, M. E. *et al.* Nanostructured ultrafast silicon-tip optical field-emitter arrays. *Nano Letters* **14**, 5035 (2014).

17. Rybka, T. *et al.* Sub-cycle optical phase control of nanotunnelling in the single-electron regime. *Nature Photonics* **10**, 667 (2016).

18. Scholes, G. D., Fleming, G. R., Olaya-Castro, A., Van Grondelle, R. Lessons from nature about solar light harvesting. *Nature Chemistry* **3**, 763 (2011).

19. Gélinas, S. *et al.* Ultrafast long-range charge separation in organic semiconductor photovoltaic diodes. *Science* **343**, 512 (2014).

20. Atwater, H. A., Polman, A. Plasmonics for improved photovoltaic devices. *Nature Materials* **9**, 205 (2010).

21. Wu, K., Chen, J., McBride, J. R., Lian, T. Efficient hot-electron transfer by a plasmon-induced interfacial charge-transfer transition. *Science* **349**, 632 (2015).

22. Sivis, M., Duwe, M., Abel, B., Ropers, C. Extreme-ultraviolet light generation in plasmonic nanostructures. *Nature Physics* **9**, 304 (2013).

23. Feist, A. *et al.* Quantum coherent optical phase modulation in an ultrafast transmission electron microscope. *Nature* **521**, 200 (2015).

24. Piazza, L. *et al.* Simultaneous observation of the quantization and the interference pattern of a plasmonic near-field. *Nature Communications* **6**, 6407 (2015).

25. Piglosiewicz, B. *et al.* Carrier-envelope phase effects on the strong-field photoemission of electrons from metallic nanostructures. *Nature Photonics* **8**, 37 (2014).

26. Lummen, T. T. *et al.* Imaging and controlling plasmonic interference fields at buried interfaces. *Nature Communications* **7**, 13156 (2016).

27. Spektor, G. *et al.* Revealing the subfemtosecond dynamics of orbital angular momentum in nanoplasmonic vortices. *Science* **355**, 1187 (2017).

28. Feist, A. *et al.* Ultrafast transmission electron microscopy using a laser-driven field emitter: Femtosecond resolution with a high coherence electron beam. *Ultramicroscopy* **176**, 63–73 (2017).

29. Raman, R. K. *et al.* Ultrafast imaging of photoelectron packets generated from graphite surface. *Applied Physics Letters* **95**, 181108 (2009).

30. Aeschlimann, M. *et al.* Adaptive subwavelength control of nano-optical fields. *Nature* **446**, 301 (2007).

31. Kubo, A. *et al.* Femtosecond imaging of surface plasmon dynamics in a nanostructured silver film. *Nano Letters* **5**, 1123 (2005).

32. Stockman, M. I. Nanofocusing of optical energy in tapered plasmonic waveguides. *Physical Review Letters* **93**, 137404 (2004).

33. Ropers, C. *et al.* Grating-Coupling of Surface Plasmons onto Metallic Tips: A Nanoconfined Light Source. *Nano Lett.* **7**, 2784–2788 (2007).

34. Schröder, B., Sivis, M., Bormann, R., Schäfer, S. & Ropers, C. An ultrafast nanotip electron gun triggered by grating-coupled surface plasmons. *Applied Physics Letters* **107**, 231105 (2015).

35. Müller, M., Kravtsov, V., Paarmann, A., Raschke, M. B. & Ernstorfer, R. Nanofocused Plasmon-Driven Sub-10 fs Electron Point Source. *ACS Photonics* **3**, 611–619 (2016).

36. Vogelsang, J. *et al.* High passive CEP stability from a few-cycle, tunable NOPA-DFG system for observation of CEP-effects in photoemission. *Optics Express* **22**, 25295 (2014).

37. Schmidt, S. *et al.* Adiabatic nanofocusing on ultrasmooth single-crystalline gold tapers creates a 10-nm-sized light source with few-cycle time resolution. *ACS Nano* **6**, 6040 (2012).

38. Quinonez, E., Handali, J. & Barwick, B. Femtosecond photoelectron point projection microscope. *Rev. Sci. Instrum.* **84**, 103710 (2013).
16

39. Bainbridge, A. R. *et al.* Femtosecond few- to single-electron point-projection microscopy for nanoscale dynamic imaging. *Structural Dynamics* **3**, 023612 (2016).

40. Longchamp, J.-N. *et al.* Imaging proteins at the single-molecule level. *Proceedings of the National Academy of Sciences* **114**, 1474 (2017).

41. Okano, Y., Hironaka, Y., Kondo, K. & Nakamura, K. G. Electron imaging of charge-separated field on a copper film induced by femtosecond laser irradiation. *Appl. Phys. Lett.* **86**, 141501 (2005).

42. Hebeisen, C. T. et al. Direct visualization of charge distributions during femtosecond laser ablation of a Si (100) surface. *Phys. Rev. B* **78**, 081403 (2008).

43. Chang, K., Murdick, R. A., Tao, Z.-S., Han, T.-R. T. & Ruan, C.-Y. Ultrafast electron diffractive voltammetry: General formalism and applications. *Modern Physics Letters B* **25**, 2099–2129 (2011).

44. Centurion, M., Reckenthaeler, P., Trushin, S. A., Krausz, F. & Fill, E. E. Picosecond electron deflectometry of optical-field ionized plasmas. *Nature Photonics* **2**, 315–318 (2008).